\documentclass[runningheads]{llncs}
\usepackage{graphicx}
\usepackage{xcolor}
\usepackage{hyperref}
\usepackage{xspace}

\newcommand{\SoBigDataITAck}{European Union - NextGenerationEU - National Recovery and Resilience Plan (Piano Nazionale di Ripresa e Resilienza, PNRR) - Project: “SoBigData.it - Strengthening the Italian RI for Social Mining and Big Data Analytics” - Prot. IR0000013 - Avviso n. 3264 del 28/12/2021\xspace}

\newcommand{\HubAck}{“ICSC – Centro Nazionale di Ricerca in High Performance Computing, Big Data and Quantum Computing", funded by European Union – NextGenerationEU\xspace}

\begin{document}
\title{A Decision Tree to Shepherd Scientists through Data Retrievability
\thanks{This work is supported by \SoBigDataITAck and by \HubAck}}
%
%
\author{Andrea Bianchi \and
Giordano d'Aloisio \and Francesca Marzi \and
Antinisca Di Marco}
\authorrunning{A. Bianchi et al.}
%
\institute{University of L'Aquila, Italy\\
\email{\{andrea.bianchi,giordano.daloisio\}@graduate.univaq.it\\
\{francesca.marzi,antinisca.dimarco\}@univaq.it
}}
%
\maketitle              
\begin{abstract}
Reproducibility is a crucial aspect of scientific research that involves the ability to independently replicate experimental results by analysing the same data or repeating the same experiment. Over the years, many works have been proposed to make the results of the experiments actually reproducible. However, very few address the importance of \textit{data reproducibility}, defined as the ability of independent researchers to retain the same dataset used as input for experimentation. Properly addressing the problem of data reproducibility is crucial because often just providing a link to the data is not enough to make the results reproducible. In fact, also proper metadata (e.g., preprocessing instruction) must be provided to make a dataset fully reproducible. 
In this work, our aim is to fill this gap by proposing a decision tree to sheperd researchers through the reproducibility of their datasets. In particular, this decision tree guides researchers through identifying if the dataset is actually reproducible and if additional metadata (i.e., additional resources needed to reproduce the data) must also be provided. This decision tree will be the foundation of a future application that will automate the data reproduction process by automatically providing the necessary metadata based on the particular context (e.g., data availability, data preprocessing, and so on). It is worth noting that, in this paper, we detail the steps to make a dataset retrievable, while we will detail other crucial aspects for reproducibility (e.g., dataset documentation) in future works.

\keywords{Data reproducibility \and FAIR principles \and Decision Tree}
\end{abstract}

\section{Introduction} 

Reproducibility is an essential aspect of scientific research that involves the ability to replicate experimental findings through independent analysis of the same data or by repeating the same experiment.
Numerous studies, particularly in the biomedical field, demonstrate the extent to which scientific findings do not pass the reproducibility test. For example, according to a Nature survey, an online questionnaire on research reproducibility administered to 1576 researchers from various disciplines showed that more than 70\% of the researchers were unable to replicate the experiments of other scientists and more than 50\% failed to reproduce their own experiments \cite{baker20161}.

This lack of reproducibility of results, known as the reproducibility crisis, significantly impacts the scientific community, making it difficult to reuse, compare, or extend the results. Furthermore, the absence of reproducible experiments not only impedes the ability of others to use our work, but also jeopardises our reputation. The reliability and validity of empirical results are fundamental scientific principles and the inability to reproduce experiments raises concerns about the credibility of the research.

Although many articles illustrate best practises for making an experiment reproducible \cite{munafo2017manifesto}, often focussing on the maintenance, publication, and documentation of algorithms \cite{fehr2016best}, very few articles focus on the importance of data reproducibility, understood as \textit{the ability of independent researchers to retain the same dataset used as input for experimentation}.
Providing a link to the raw dataset alone is often insufficient to make the results reproducible \cite{pawlik2019link}. Typically, datasets are manipulated in the preprocessing phase to be tailored to the experiment's needs, and this stage can significantly impact the outcomes. Consequently, in addition to the raw dataset, it is also necessary to provide the preprocessing script that was used to generate the final dataset used as input for the algorithm. Moreover, data are often protected by regulations (e.g., in the medical or public field) and can not be publicly shared, hence different methods to access them (if possible) must be provided. 

This paper aims to construct a decision tree that guides researchers through the reproducibility of their datasets. The proposed decision tree provides a structured guide that not only allows researchers to verify whether their datasets are reproducible, but also outlines the clear steps to be taken to ensure reproducibility. By following the decision tree, researchers can gain an understanding of the factors that contribute to the reproducibility of their datasets and take the appropriate actions to address potential problems. 

It is worth noting that in this work we consider a dataset \textit{reproducible} if it meets two criteria: it is \textit{retrievable} and it is \textit{usable}. For \textit{retrievability} we mean the possibility for other researchers to obtain the dataset in the same format as it was used in the original experiment. For \textit{usability} we mean the presence of a proper documentation that describes all the information to use the dataset properly (e.g., describes the meaning of the columns, the purpose of the dataset, and so on). In this paper, we focus on the first aspect of reproducibility (i.e., retrievability), while we will address usability in future work. 

The paper is organised as follows. We motivate the discussion on data reproducibility based on our experiences and related work in Section \ref{sec:related}. In Section \ref{sec:tree}, we introduce our decision tree. Conclusions wrap up the paper in Section \ref{sec:conclusion}.

\section{Motivation and Related Works}\label{sec:related}

In this section, we first describe the motivation that led us to define the decision tree described in Section \ref{sec:tree}. Next, we discuss some related work in the context of data reproducibility.

The problem of data reproducibility has gained relevance in the literature in recent years. In \cite{miyakawa_no_2020} the author highlights how the absence of raw data may be one of the causes of the so-called \textit{reproducibility crisis}. However, in \cite{pawlik2019link} the authors also highlight that even if the raw data are accessible, providing just a link to them is not enough. In fact, metadata (i.e., additional information needed to use the data correctly) must also be provided to make an experiment reproducible. For example, data may need to be preprocessed, so preprocessing instructions must also be provided. Nevertheless, metadata are context-dependent and different metadata must be provided based on, for instance, the type of data used or the level of accessibility. For instance, in the medical field, data are often protected by regulations and can not be publicly shared. This implies that different methods to use the data (e.g., accessing them locally using a machine not connected to the web) must be provided. Similarly, data from the education field are often accessible to use, but can not be publicly shared in raw format, but only in preprocessed format. Finally, data used in the more general data science domain are often publicly available and shareable.

This heterogeneity in the metadata needed for the reproducibility of data is also confirmed by two of our previous works. In \cite{daloisio_debiaser_2023} we used an open access data set for our experiments. In particular, these datasets were taken from open access repositories and were properly documented in their relative papers. In addition, we performed some preprocessing steps to prepare the data for our experiments. Hence, in this case, since the data were open access and fully distributable, to make them reproducible, we provided directly a link to the preprocessed data and a reference to papers first describing them. 
Differently, in \cite{cancers13194863} the data used for the experiments were not open access, and therefore a direct link to the data could not be provided. In an experimental study conducted in the medical domain, the data were initially inaccessible due to privacy concerns. In order to gain access to the data, a formal collaboration was established with the entity that had collected the dataset. Through this collaboration, the entity provided us with all the necessary information about the dataset, including details about all instances and the preprocessing steps that had been applied to the raw data. This allowed us to generate a preprocessed version of the dataset that could be used for our experiment, enabling us to proceed with our research.
These two experiences describe well the different possibilities for making data reproducible. 

Our aim is to take a first step in solving this problem by providing a decision tree that guides researchers through the reproducibility of their datasets. In particular, in this work, we focus on data retrievability, while we will detail other fundamental reproducibility aspects (i.e., usability) in future work. 

In terms of related work, few studies have explored the problem of data reproducibility. Gebru et al., in \cite{gebru_datasheets_2021} presented a datasheet to document datasets and facilitate the communication between datasets creators and consumers. In particular, they provide a set of questions useful to document a dataset and make it reproducible. In \cite{monks_strengthening_2019} Monk et al. introduced the STRESS guidelines, a checklist to make the empirical simulation results reproducible. The checklist is divided into six sections, namely \textit{objectives}, \textit{logic}, \textit{data}, \textit{experimentation}, \textit{implementation}, and \textit{code access}. In our work, our aim is to automate some parts of the data documentation process depicted in these papers by proposing a decision tree that actually guides the researcher through the reproducibility of the dataset. 
In \cite{ning_easy_2020}, Ning et al. proposed CROWDAQ, a tool that standardises the data collection pipeline in the context of machine learning with customisable user interface components, automated annotator qualification, and saved pipelines in a re-usable format. The authors show how their tool can help standardising the data collection pipeline and making it reproducible. However, they do not consider several data reproducibility contexts, like, for instance, where the data used is not open access. Furthermore, they do not consider additional metadata required for data reproducibility (e.g., data documentation).

\section{Decision Tree}\label{sec:tree}
In order to ensure data reproducibility, it is important to have a clear understanding of the dataset's accessibility and processing methods, if any. 
As already mentioned in the sections above, in this work we focus on the retrievability of the dataset (i.e., making it accessible to other researchers), while we will detail the usability of the dataset in future work. In this section, we use a decision tree to formulate a schema for examining the retrievability of the dataset, focusing on its accessibility for future research and analysis.

\begin{figure}
    \centering
    \includegraphics[width=\textheight,angle=90]{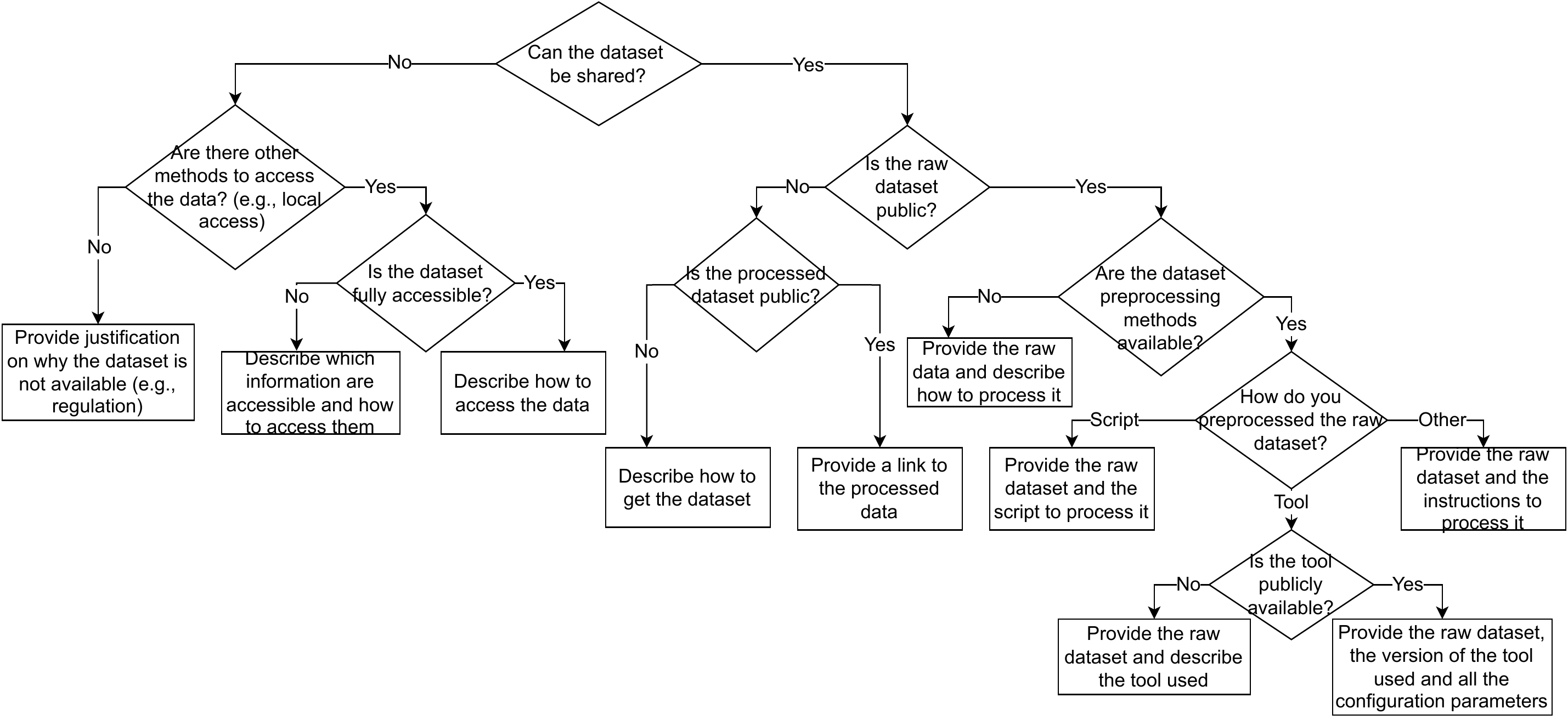}
    \caption{Data Retrievability Tree}
    \label{fig:tree}
\end{figure}

The decision tree is depicted in figure \ref{fig:tree} and is also available on the Zenodo repository \cite{andrea_bianchi_2023_7685205}. The decision tree is composed of eight main questions, represented as internal diamond nodes. Leaf are represented by squared nodes and give specific suggestions about what to provide in terms of metadata or material. Each set of answers leads to a specific leaf of the tree.
The tree structure allows for a clear and structured approach to handling different scenarios when sharing datasets, by asking relevant questions and providing specific suggestions at each decision point.

The tree can be divided into two macro-areas, which are based on the fact that the dataset can be shared (i.e., make it publicly available) or not. If the dataset can not be shared, then we move to the left part of figure \ref{fig:tree} asking if there are other methods to access the data (e.g., the dataset is locally available inside a machine not connected to the web). If there are no methods to access the data, then the dataset is not reproducible and the researcher must provide a reason on why it is not reproducible (i.e., regulations). If instead there are other methods to access the data and the data is also fully accessible (i.e, all the information of the dataset can be viewed), then the researcher must describe how to access the data. If not all the information of the dataset is accessible (which is very common, for instance, in medical records where local, national, and/or international regulations may restrict certain data \cite{cancers13194863}), then the researcher must also describe which information are accessible and how they are accessible.

Instead, the right part of the tree in Figure \ref{fig:tree} is reached when the dataset can be shared. In particular, the dataset can be shared in two forms: the raw (unprocessed) dataset and the final preprocessed dataset. A first decision branch is made on the question if the raw dataset is public or not.
If the raw dataset is not public, but the preprocessed dataset is, then the researcher must provide the source link to the preprocessed dataset. If instead either the raw or the preprocessed data are not public, then the researcher must describe how to get the raw and/or the preprocessed dataset (e.g., by providing a formal request to a platform).

If the raw dataset is public, then a new decision branch is made on the question if the dataset preprocessing methods are available or not. If the preprocessing methods are not available, then the researcher must provide the source to the raw dataset and describe how to process it (if preprocessing techniques were applied). If preprocessing methods are instead available, then another decision branch is made on the question on how the dataset was preprocessed. If the dataset was preprocessed using a script, then the script must be provided together with the raw dataset. If a tool was used to preprocess the data and the tool is publicly available, then the researcher must provide the raw dataset and the tool along with the tool version used and its configuration parameters if present. If instead a tool was used, but it is not publicly available, then the researcher must provide the raw datased and describe which tool was used along with its version and configuration parameters if present. Finally, if other methods (different from a script or a tool) were used, then the researcher must provide the raw dataset and the instructions on how to process it.

The decision tree presented in this section can serve as a helpful tool to guide researchers through the reproducibility of their datasets, thus contributing to the transparency, and the replicability of their research and for that to the advancement of scientific knowledge.  




\section{Conclusions and Future Work}\label{sec:conclusion}
In this paper, we presented a first version of a decision tree to guide researchers through assessing and verifying the reproducibility of datasets. In particular, here we focused on the steps needed to make a dataset retrievable. We first described the motivations of our approach and then detailed the depicted decision tree.  

In future work, we plan to first extend the decision tree to other fundamental aspects of reproducibility (e.g., documentation). Next, the final tree will be the foundation of a web-based service that will be designed to generate the metadata necessary to improve the reproducibility of the datasets. The service will function by administering a questionnaire, which helps identify critical factors that affect reproducibility. The service will generate the required metadata based on the questionnaire responses, enabling the dataset to be reproduced efficiently. 

\bibliographystyle{splncs04}
\bibliography{bibliography}

\end{document}